\documentclass[10pt,letterpaper,twocolumn]{revtex4}
\usepackage{graphicx}

\begin{document}


\title{Fano resonance in quadratic waveguide arrays}

\author{Andrey E. Miroshnichenko and Yuri S. Kivshar}

\address{Nonlinear Physics Centre and Centre for Ultra-high
bandwidth Devices for Optical Systems (CUDOS), Research School of
Physical Sciences and Engineering, Australian National University,
Canberra ACT 0200, Australia}

\author{Rodrigo A. Vicencio and Mario I.  Molina}

\address{Departamento de F\'{\i}sica, Facultad de Ciencias,
Universidad de Chile, Casilla 653, Santiago, Chile}

\begin{abstract}
We study resonant light scattering in arrays of channel optical
waveguides where tunable quadratic nonlinearity is introduced as
nonlinear defects by periodic poling of single (or several)
waveguides in the array. We describe novel features of wave
scattering that can be observed in this structure and show that it
is a good candidate for the first observation of Fano resonance in
nonlinear optics.
\end{abstract}

\maketitle

The study of nonlinear dynamics and spatial solitons in optical
systems has recently attracted a great deal of
attention~\cite{book}. In particular, many specific properties of
nonlinear lattice systems can be analyzed for arrays of weakly
coupled optical waveguides where both nonlinearity and diffraction
may differ dramatically compared to those in the corresponding
continuous systems~\cite{review,review2}.

During last years a growing interest is observed in the study of
nonlinear optics associated with the so-called {\em quadratic
nonlinearities} which may produce the effects resembling those
known to occur in cubic nonlinear materials. Typical examples are
all-optical switching phenomena in interferometric or coupler
configurations as well as the formation of spatial and temporal
solitons in planar waveguides~\cite{buryak}. Recently, it was
demonstrated experimentally~\cite{exp_chi2} that arrays of coupled
channel waveguides fabricated in a periodically poled Lithium
Niobate slab represent a convenient system to verify
experimentally many theoretical predictions, including the first
observation of two-frequency discrete solitons mutually locked by
quadratic nonlinearity. These experimental observations open many
perspectives for employing larger nonlinearities in lattice
systems made of quadratic materials.

In this Letter we suggest to employ the arrays of weakly coupled
nonlinear quadratic waveguides for the study of novel effects in
resonant light scattering. In particular, we show that when
periodic poling is applied to just a few waveguides in the array,
it creates a nonlinear defect~\cite{defect,chi2_defect} that
possesses specific resonant scattering properties and may be
employed for the first experimental observation of Fano resonance
in nonlinear optics.

\begin{figure}[tbp]
\includegraphics[width=8.4cm]{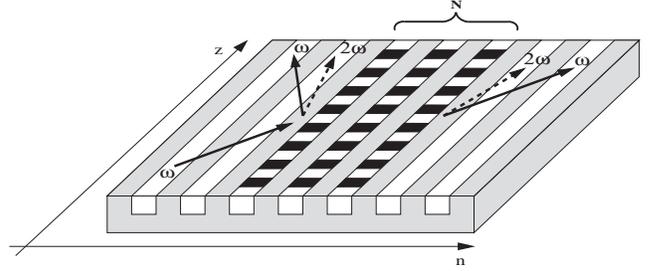}
\caption{Schematic of an array of channel waveguides with
quadratic nonlinearity where nonlinear defects are created by
periodic poling. Arrows show the scattering process.}
\label{Fig1}
\end{figure}

Following the waveguide design recently implemented in
Ref.~\cite{exp_chi2} for the observation of discrete quadratic
solitons, we consider a discrete model describing an array of
weakly coupled linear waveguides where one or several neighboring
waveguides have periodic poling and therefore possess a quadratic
nonlinear response (see Fig.~\ref{Fig1}). When the matching
conditions are satisfied, the fundamental-frequency (FF) mode with
the frequency $\omega$ can generate parametrically the
second-harmonic (SH) wave at the frequency $2\omega$, so that such
a structure with several poled waveguides may behave as {\em a
nonlinear defect} with localized quadratic nonlinearity. The
continuous version of this model has been studied
earlier~\cite{sukh_chi2}.

In the tight-binding approximation usually employed in the theory
of discrete lattices~\cite{review2}, the effective equations for
the complex envelopes of the FF wave ($u_n$) and its SH component
($v_n$) coupled at the defect waveguides with $n=0, \ldots, N$ can
be written in the dimensionless form
\begin{equation}  \label{eq:1}
   \begin{array}{l} {\displaystyle
      \mathrm{i} {du_{n}\over{d z}} + c_{u}( u_{n+1} + u_{n-1} ) + 2
\sum_{m=0}^N u_{m}^{*} v_{m} \delta_{n m} =0,
   } \\*[9pt] {\displaystyle
      \mathrm{i} {dv_{n}\over{d z}} + c_{v}( v_{n+1} + v_{n-1} ) -\Delta v_{n}
+ \sum_{m=0}^N u_{m}^{2} \delta_{n m}=0,
   } \end{array}
   \label{eq_1}
\end{equation}
where $c_u$ and $c_v$ are the coupling coefficients, $\delta_{nm}$
is the Kronecker symbol and $\Delta$
is the phase mismatch parameter assumed to be identical for all
waveguides.

First, we analyze the scattering of a plane FF wave at the
frequency $\omega$ by the a single ($n=0$) quadratic defect
waveguide (`impurity site'). After the interaction with the
quadratic waveguide, the FF wave generates a SH wave which could
either propagate or get trapped being guided by the defect
waveguide. To calculate the transmission coefficient  $t(k)$ of
the FF wave, we present the fields in the form,
\begin{equation}
u_{n}(z) = \mathrm{exp}(\mathrm{i}\beta_1 z)\left\{ \begin{array}{ll}
    I \mathrm{exp}(\mathrm{i} k n) + R \mathrm{exp}(-\mathrm{i} k n) & \mbox{$n < 0$}\\
    T \mathrm{exp}(\mathrm{i} k n)  & \mbox{$n \geq N$},
    \end{array}
        \right.
\end{equation}
\begin{equation}
v_{n}(z) = \mathrm{exp}(\mathrm{i}\beta_2 z) \left\{ \begin{array}{ll}
    {\tilde{R}} \mathrm{exp}(-\mathrm{i} q n) & \mbox{$n < 0$}\\
    {\tilde{T}} \mathrm{exp}(\mathrm{i} q n) & \mbox{$n \geq N$},
    \end{array}
        \right.
\end{equation}
where $\beta_1=2c_u\cos k$ and $\beta_2=2c_v\cos q-\Delta$  
are propagation constants of the FF and SH respectively,
and $k$ and $q$ are corresponding transverse wavenumbers.
By using the phase-matching condition \cite{sukh_chi2} ($2\beta_1=\beta_2$) 
we obtain the relation $c_v \cos q - (\Delta/2) = 2c_u \cos k$, 
that defines the dependence $q=q(k)$.
For $k_{\rm min} < k < k_{\rm max}$, the function $q(k)$ takes
real and positive values. Outside this interval, the values of
$q(k)$ are purely imaginary, and they correspond to localized
(non-radiating) states trapped by the defect waveguide at $n=0$. A
simple calculation yields the result: $k_{\rm max,min} =
\cos^{-1}[(c_{v} \pm \Delta/2)/2 c_{u}]$, when these values are real
and positive or  zero, otherwise. 

Evaluating the mode coupling at the impurity ($n = 0$) and
neighboring ($n=-1$) sites allows to obtain the relations, $T=I+R$
and ${\tilde{R}}= {\tilde{T}}$, and derive a nonlinear equation
for the transmission coefficient $t=|T|^2/|I|^2$ in the form,
$t[1+b(k)t]^2 = 1$, where $b(k) = |I|^2\left[2c_{u}c_{v}\sin k\sin
q(k)\right]^{-1}$, that has only one real solution. The resonant
scattering, when a localized SH field is generated, can be
analyzed similarly by replacing $q \rightarrow iq$.

\begin{figure}[tbp]
\includegraphics[width=8.4cm]{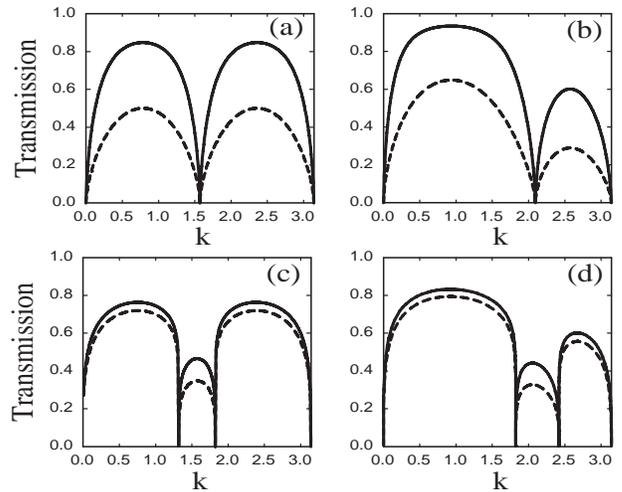}
\caption{Examples of the FF transmission
coefficient for the scattering by the quadratic
impurity waveguide, for $c_u=1$, and (a) $c_v=0$, $\Delta=0$,
(b)  $c_v=0$, $\Delta=2$, (c)  $c_v=0.5$, $\Delta=0$, and
(d)  $c_v=0.5$, $\Delta=2$, for two values of the intensity:
$I=1$ (solid line) and $I = 2$ (dashed lines).}
\label{Fig2}
\end{figure}

The study of wave scattering in this system predicts the resonant
suppression of transmission at some points, i.e. $t(k_{\rm
min, max})=0$ (see Fig.~\ref{Fig2}). 
We demonstrate below that these resonant
reflections correspond to {\em a novel type} of the well-known
Fano resonance~\cite{fano}. Indeed, according to the Fano
theory~\cite{fano}, destructive interference and resonant
suppression of transmission are observed when there exists {\em a
localized state} coupled to the propagating channel with the
energy inside the linear spectrum. Note, that $q(k_{\rm min})=0$
and $q(k_{\rm max})=\pi$, i.e. these values define the band edges
of the propagation spectrum of the propagating SH field, and the
resonances take place when the SH field is generated. This
situation seems to be in contradiction with the classical
definition of the Fano resonance. However, below we demonstrate in
more details that this kind of the resonant scattering can be
indeed defined as being associated with the Fano resonance.

First, we consider the simplest case when the coupling between the
SH modes of different waveguides vanishes, i.e. $c_v=0$, which is
indeed the case of the recent experiments~\cite{exp_chi2}. Then,
in the stationary case, the coupled equations (\ref{eq:1}) can be
written in the form 
\begin{equation}  \label{eq36}
   \begin{array}{l} {\displaystyle
    \beta_1 u_n = c_u(u_{n+1}+u_{n-1})+2u_0^*v\delta_{n0},
   } \\*[9pt] {\displaystyle
2\beta_1 v = -\Delta v+u^2_0,
   } \end{array}
\end{equation}
This model (\ref{eq36}) describes the main propagation channel for
the field $u_n$ and an additional discrete mode $v=v_0$ coupled to
it parametrically, and it is similar to the so-called
Fano-Anderson model~\cite{mahan}, except that the coupling here is
{\em nonlinear}, that makes the scattering problem nonlinear.

To simplify the analysis, we eliminate the discrete mode described by
the second
equation and obtain,
\begin{eqnarray}
\beta_1 u_n =
c_u(u_{n+1}+u_{n-1})+\frac{2|u_0|^2u_0}{2\beta_1+\Delta}\delta_{n0},
\label{eq38}
\end{eqnarray}
which is an effective equation for the propagation channel that
contains a scattering potential. The strength of this {\em nonlinear
resonant scattering potential} depends on the incoming intensity $|I|^2$
and propagation constant $\beta_1(k)$.
If $\Delta$ is chosen such that $k_F$ is between $0$ and
$\pi$, and
\begin{eqnarray}
2\beta_1(k_F)=-\Delta,
\label{eq39}
\end{eqnarray}
our potential becomes infinitely large for a particular frequency
$\beta_1(k_F)$, which will lead to the perfect reflection. Note
here, that $k_F=k_{\rm min, max}$ for our case $c_v=0$.
Indeed, after some algebra we can write down the equation for the
transmission coefficient in the following form
\begin{eqnarray}
t^3+\gamma(k)(t-1)=0,
\label{eq40}
\end{eqnarray}
where $\gamma(k)=(4c_u\cos k+\Delta)^2c_u^2\sin^2k/|I|^4$. From
this equation one can see that, when $\gamma(k)=0$ then
transmission coefficient $t$ vanishes. This happens at the
wavenumbers $k=0$ and $k=\pi$, which correspond to the band edges of
the propagation spectrum, and also at $k=k_F$. In the latter case,
this is
exactly the Fano resonance.

When the coupling between the SH modes in the waveguide array does
not vanish (i.e. $c_v \not=0$), it leads to the appearance of the
spectrum of propagating SH modes, $\beta_2[q(k)]$. At the band edges
of this spectrum, $\beta_2(0)$ and $\beta_2(\pi)$, which
correspond to $k=k_{\rm min}$ and $k=k_{\rm max}$, the propagating
SH
field is described as a standing constant-amplitude mode of the forms
$v_n=v_0$
and $v_n=(-1)^nv_0$, respectively, where $v_0$ is constant.
Therefore, for these two cases we can again obtain the single-site
equation for the second scattering channel. By applying similar approach
to these particular cases, we obtain two conditions for the Fano
resonance,
\begin{eqnarray}
2\beta_1(k_{F_{1,2}})=-\Delta\pm 2c_v,
\label{eq41}
\end{eqnarray}
which occur exactly when the propagation constant $2\beta$ of the generated SH 
field
coincides with either the propagation constant $\beta_2(0)$ or $\beta_2(\pi)$,
i.e. at the band edges of the linear spectrum of the propagating SH
modes. In other words, in such situations we excite
'constant modes' by a local perturbation. Since the group velocity of
these modes vanishes at the band edges, any local excitation could not
propagate at the given frequency. It makes these modes 
\textit{effectively
local} and leads, finally, to the phenomenon of Fano resonance.

\begin{figure}[tbp]
\includegraphics[width=8.4cm,height=7cm]{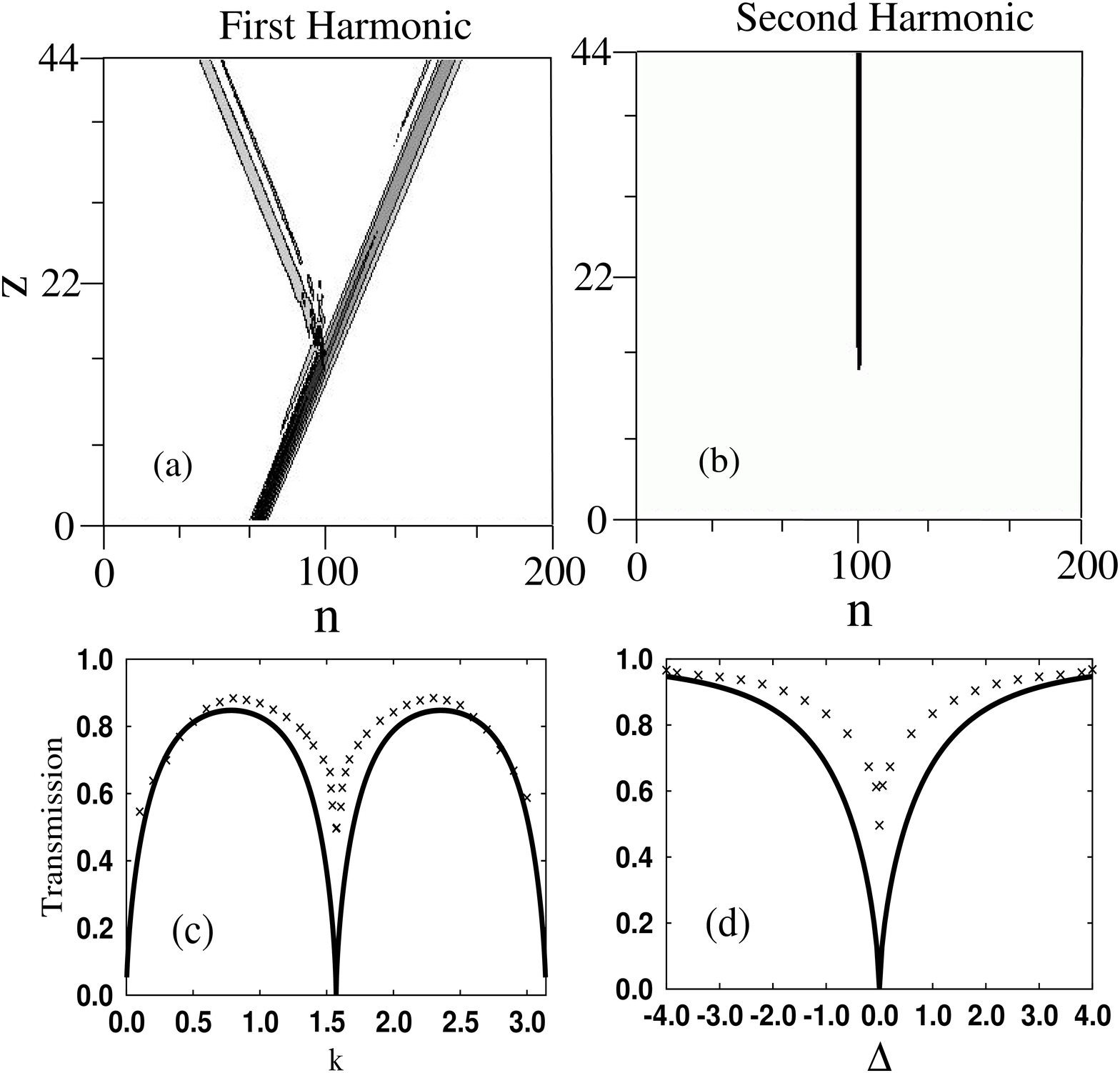}
\caption{Top: Resonant generation of the trapped second-harmonic field
by a Gaussian beam of the FF (a) and SH (b) fields for the parameters $c_u=1$, 
$c_v=0$,
$\Delta=0$, and $k=\pi/2$. Bottom: Comparison of the transmission
coefficients of the plane waves (solid) and Gaussian beam (crosses)
for $c_u=1$, $c_v=0$ and (c) $\Delta=0$ and (d) $k=\pi/2$.
Transmission coefficient of the Gaussian
beam does not vanish due to its finite spectral width. }
\label{Fig3}
\end{figure}

We note here that, in our physical system of quadratic waveguides,
the coupling between the first and second propagation channels is
nonlinear and, therefore, it depends on the intensity of the
incoming wave. According to the formulas (\ref{eq39}) and
(\ref{eq41}) the position of Fano resonances does not depend on
the value of this coupling due to its local nature~\cite{flach}.
As a consequence, this novel type of Fano resonance should exist
for any intensity of incoming waves similar to the conventional
Fano resonance in the linear theory. But the width of the resonance
depends on this coupling\cite{aem} and, therefore, on incoming intensity 
of light
(see Fig.~\ref{Fig2}).

In order to check the validity of our plane wave analysis and the 
manifestation
of the effect in a realistic experiment, we have performed the numerical 
simulation of
the Gaussian beam scattering. The results are summarized in
Fig.~\ref{Fig3}, and they are in a
good agreement with the theory of plane wave scattering.

In the case of $N$ defects and vanishing coupling between them
($c_v=0$), Fano resonance does not change its position
(\ref{eq39}) \cite{periodic}. After scattering by the first defect close to Fano
resonance, other waveguides become almost transparent due to a
small incoming intensity. Therefore, the width of the resonance
will remain almost the same as in the case of a single defect.


In conclusion, we have analyzed a novel waveguide structure where
an optical analog of the Fano resonance can be observed as
peculiarities of the resonant scattering and second-harmonic
generation in the quadratic waveguide arrays. We believe this kind
of waveguide structures, already fabricated for the
observation of discrete quadratic solitons~\cite{exp_chi2}, is
a good candidate for the first experimental observation of Fano
resonances in nonlinear optics.

This work was partially supported by the Australian Research
Council, a doctoral fellowship from Conicyt, and Fondecyt grants
1020139 and 7020139 in Chile. The authors thank Andrey Sukhorukov
for useful discussions.

\end{document}